**Title: "Who's Winning? Clarifying Estimands Based on Win Statistics in Cluster Randomized Trials"**

Short Title: "Win Statistics in CRTs"

Author List: Kenneth M. Lee,[1,2,3*] Xi Fang,[4] Fan Li,[4,5] Michael O. Harhay[1,2,3,6]

[1] Department of Biostatistics, Epidemiology and Informatics, University of Pennsylvania, Philadelphia, PA, USA

[2] Center for Clinical Trials Innovation, Department of Biostatistics, Epidemiology, Informatics, University of Pennsylvania, Philadelphia, PA, USA

[3] Palliative and Advanced Illness Research Center, Perelman School of Medicine, University of Pennsylvania, Philadelphia, PA, USA

[4] Department of Biostatistics, Yale School of Public Health, New Haven, CT, USA

[5] Center for Methods in Implementation and Prevention Science, Yale School of Public Health, New Haven, CT, USA

[6] MRC Clinical Trials Unit, University College London, London, UK

*Corresponding author. Center for Clinical Epidemiology and Biostatistics, University of Pennsylvania School of Medicine, 3600 Civic Center Boulevard, Philadelphia, PA 19104

E-mail: kenneth.lee@pennmedicine.upenn.edu.

## Abstract

Treatment effect estimands based on win statistics, including the win ratio, win odds, and win difference are increasingly popular targets for summarizing endpoints in clinical trials. Such win estimands offer an intuitive approach for prioritizing outcomes by clinical importance. The implementation and interpretation of win estimands is complicated in cluster randomized trials (CRTs), where researchers can target fundamentally different estimands on the individual-level or cluster-level. We numerically demonstrate that individual-pair and cluster-pair win estimands can substantially differ when cluster size is informative: where outcomes and/or treatment effects depend on cluster size. With such informative cluster sizes, individual-pair and cluster-pair win estimands can even yield opposite conclusions regarding treatment benefit. We describe consistent estimators for individual-pair and cluster-pair win estimands and propose a leave-one-cluster-out jackknife variance estimator for inference. Despite being consistent, our simulations highlight that some caution is needed when implementing individual-pair win estimators due to finite-sample bias. In contrast, cluster-pair win estimators are unbiased for their respective targets. Altogether, careful specification of the target estimand is essential when applying win estimators in CRTs. Failure to clearly define whether individual-pair or cluster-pair win estimands are of primary interest may result in answering a dramatically different question than intended.

# Introduction

Different treatment effect estimands based on win statistics, including the win ratio, win odds, and win difference, have recently emerged as attractive targets for comparing complex endpoints in clinical trials. These "win estimands" offer an intuitive framework for prioritizing endpoints by clinical importance by comparing these endpoints between every treated and control patient.[1–3] Simply, in each pair of treated and control patients, a patient "wins" if they have the better outcome on the highest-priority endpoint.

Among different randomized controlled trial designs, the cluster randomized trial (CRT) has become increasingly popular. CRTs involve randomizing groups of individuals (e.g., hospitals, schools, or communities) to different intervention arms.[4] These designs are commonly employed when interventions operate at the cluster level, where contamination between treatment arms may be a concern, or when individual randomization is logistically impractical.

Despite the growing popularity of both CRTs and win estimands, less attention has been devoted to the appropriate definition and estimation of win estimands in clustered settings. As previously highlighted in the context of typical average treatment effects and some other ratio estimands (i.e., odds ratio, risk ratio), different analytical approaches in CRTs can target fundamentally different estimands on either the individual-level or cluster-level.[5–7] The individual-average estimand asks, "*what is the expected change in outcome associated with treatment among the population of individuals, regardless of cluster membership*?" The cluster-average estimand asks, "*what is the expected change in outcome associated with treatment among the population of clusters?*" The appropriate choice of estimand then depends on the scientific question of interest, with these different estimands offering fundamentally different interpretations.[5–7] This distinction becomes particularly important when cluster size is informative and outcomes and/or treatment effects are associated with the number of individuals in each cluster. In the presence of such informative cluster sizes, individual-level and cluster-level estimands have been observed to dramatically differ in magnitude.[5,6]

In this article, we extend this discussion of estimands in CRTs to clarify the distinction between individual-pair and cluster-pair win estimands in CRTs under the potential outcomes framework. We introduce consistent estimators for each target estimand, propose variance estimation via convenient jackknife methods, and demonstrate through worked examples and simulations that these two estimands can substantially diverge in the presence of informative cluster sizes. Ultimately, we hope to provide investigators with conceptual insights and methodological tools to ensure that their implemented win estimators address the intended scientific question in increasingly complex CRT contexts.

# Methods

## Win Estimands in Individual-Randomized Trials

We begin by reviewing win estimands in the standard (non-clustered) individual-randomized trial setting with the potential outcomes framework. Consider $N$ individuals indexed by $i = 1,\ldots,N$, where $Z_i$ denotes treatment assignment ($Z_i = 1$ for treatment, $= 0$ for control) and $Y_i$ denotes the observed outcome.[8] Under the Stable Unit Treatment Value Assumption (SUTVA), the observed outcome relates to potential outcomes as $Y_i = Z_i Y_i(1) + (1 - Z_i)Y_i(0)$, where $Y_i(1)$ and $Y_i(0)$ are the individual potential outcomes under treatment and control, respectively.

In individual-randomized trials, the basic win probability estimator compares outcomes from all treated individuals $i$ to all control individuals $j$, and can be written as:

$$\hat{\tau}_{win} = \frac{\sum_{i=1}^{N} \sum_{j=1}^{N} Z_i(1 - Z_j) \text{I}(Y_i > Y_j)}{\sum_{i=1}^{N} \sum_{j=1}^{N} Z_i(1 - Z_j)}$$

where $i, j \in 1, \ldots, N$. In this setting and under SUTVA, this estimator converges in probability to the following win probability estimand:

$$\tau_{win} = P(Y_i(1) > Y_j(0)).$$

Here, $\tau_{win}$ represents the probability that a randomly selected treated potential outcome $i$ exceeds a randomly selected control potential outcome $j$. We can then similarly define $\tau_{loss} = P(Y_i(1) < Y_j(0))$ and $\tau_{tie} = P(Y_i(1) = Y_j(0))$. Overall, these win, loss, and tie probability estimands are taken as averages over the marginal distributions of the potential outcomes $Y(1)$ and $Y(0)$.[9,10] More discussion of the estimand definitions with potential outcomes are included in Appendix S1.

Based on the above component estimands, the final summary win estimands ($WE$): win-ratio, win-odds, and win-differences, can then be defined as:

$$WR = \frac{\tau_{win}}{\tau_{loss}}, \quad WO = \frac{\tau_{win} + 0.5\tau_{tie}}{\tau_{loss} + 0.5\tau_{tie}}, \quad WD = \tau_{win} - \tau_{loss}.$$

The popular win ratio ($WR$) is the probability that treatment wins over the probability that control wins[3]; an estimand value > 1 reveals a favorable treatment effect. For example, $WR = 2$ indicates that among all possible untied pairs, the treatment arm wins 2 out of 3 times. The win odds ($WO$) extends the win ratio to consider ties and may be preferable when such ties exist.[1] Finally, the win difference ($WD$, also referred to as "net benefit") is the difference between the win and loss probabilities.[2]

## Win Estimands in Cluster-Randomized Trials

In a typical parallel cluster randomized trial (CRT), we have individuals $k = 1, \ldots, N_i$ nested within clusters $i = 1, \ldots, M$. Treatment is assigned at the cluster level, so $Z_i$ now indicates whether cluster $i$ is assigned to treatment. Under the cluster-level SUTVA, the observed outcome for individual $k$ in cluster $i$ relates to potential composite outcomes as $Y_{ik} = Z_i Y_{ik}(1) + (1 - Z_i) Y_{ik}(0)$, where $Y_{ik}(1)$ and $Y_{ik}(0)$ are the individual-level potential outcomes under treatment and control, respectively. Extending the CRT estimands framework,[5,6] we can then define two distinct types of win estimands: the individual-pair and the cluster-pair win estimands.

## Individual-Pair Estimators & Estimands

The individual-pair estimand treats each individual as the unit of inference, giving equal weight to all participants regardless of which cluster they belong to. If researchers are primarily interested in an effect on the average individual across the population of participants, an individual-pair estimand may be more appropriate. This may be suitable with participant-level outcome measures such as mortality, quality of life, etc.

The individual-pair win probability estimator is then:

$$\hat{\tau}_{win,ind} = \frac{\sum_{i=1}^{M} \sum_{j=1}^{M} Z_i(1-Z_j) \sum_{k=1}^{N_i} \sum_{l=1}^{N_j} I(Y_{ik} > Y_{jl})}{\sum_{i=1}^{M} \sum_{j=1}^{M} Z_i(1-Z_j) N_i N_j}.$$

This estimator compares all individuals across treated clusters to all individuals across control clusters, with each individual-pair comparison receiving equal weight. Under cluster-level SUTVA, the individual-pair win probability estimator converges in probability to the following individual-pair win probability estimand with potential outcomes[11]:

$$\tau_{win,ind} = \frac{E\left[\sum_{k=1}^{N_i} \sum_{l=1}^{N_j} I(Y_{ik}(1) > Y_{jl}(0))\right]}{E[N_i N_j]}$$

where the expectations are over all pairs clusters $i$ and $j$. The individual-pair loss and tie probability estimators, $\hat{\tau}_{loss,ind}$ and $\hat{\tau}_{tie,ind}$, and estimands, $\tau_{loss,ind}$ and $\tau_{tie,ind}$, are analogously defined.

Altogether, we can then define the individual-pair win ratio, win odds, and win difference estimators:

$$\widehat{WR}_{ind} = \frac{\hat{\tau}_{win,ind}}{\hat{\tau}_{loss,ind}}, \quad \widehat{WO}_{ind} = \frac{\hat{\tau}_{win,ind} + 0.5\hat{\tau}_{tie,ind}}{\hat{\tau}_{loss,ind} + 0.5\hat{\tau}_{tie,ind}}, \quad \widehat{WD}_{ind} = \hat{\tau}_{win,ind} - \hat{\tau}_{loss,ind}$$

and estimands:

$$WR_{ind} = \frac{\tau_{win,ind}}{\tau_{loss,ind}}, \quad WO_{ind} = \frac{\tau_{win,ind} + 0.5\tau_{tie,ind}}{\tau_{loss,ind} + 0.5\tau_{tie,ind}}, \quad WD_{ind} = \tau_{win,ind} - \tau_{loss,ind}.$$

## Cluster-Pair Estimators & Estimands

The cluster-pair estimand treats each cluster as the unit of inference, giving equal weight to all clusters regardless of their size. This can be achieved by applying inverse cluster-size weights to the previously described individual-pair comparisons. If researchers are primarily interested in an effect on the average cluster across the population of clusters, a cluster-pair estimand may be more appropriate. This may be suitable when outcomes capture "cluster behavior", such as prescribing habits for a behavior change intervention.

The cluster-pair win probability estimator is then:

$$\hat{\tau}_{win,clus} = \frac{\sum_{i=1}^{M} \sum_{j=1}^{M} Z_i(1-Z_j) \frac{1}{N_i N_j} \sum_{k=1}^{N_i} \sum_{l=1}^{N_j} I(Y_{ik} > Y_{jl})}{\sum_{i=1}^{M} \sum_{j=1}^{M} Z_i(1-Z_j)}.$$

This estimator first computes the average win proportion within each pair of clusters, then averages across cluster pairs. With cluster-level SUTVA, the cluster-pair win probability estimator converges in probability to the following cluster-pair win probability estimand with potential outcomes[11]:

$$\tau_{win,clus} = E\left[\frac{1}{N_i N_j}\sum_{k=1}^{N_i}\sum_{l=1}^{N_j} \mathrm{I}(Y_{ik}(1) > Y_{jl}(0))\right].$$

The cluster-pair loss and tie probability estimators, $\hat{\tau}_{loss,clus}$ and $\hat{\tau}_{tie,clus}$, and estimands, $\tau_{loss,clus}$ and $\tau_{tie,clus}$, are analogously defined.

Altogether, we can then define the cluster-pair win ratio, win odds, and win difference estimators:

$$\widehat{WR}_{clus} = \frac{\hat{\tau}_{win,clus}}{\hat{\tau}_{loss,clus}}, \qquad \widehat{WO}_{clus} = \frac{\hat{\tau}_{win,clus} + 0.5\hat{\tau}_{tie,clus}}{\hat{\tau}_{loss,clus} + 0.5\hat{\tau}_{tie,clus}}, \qquad \widehat{WD}_{clus} = \hat{\tau}_{win,clus} - \hat{\tau}_{loss,clus}$$

and estimands:

$$WR_{clus} = \frac{\tau_{win,clus}}{\tau_{loss,clus}}, \qquad WO_{clus} = \frac{\tau_{win,clus} + 0.5\tau_{tie,clus}}{\tau_{loss,clus} + 0.5\tau_{tie,clus}}, \qquad WD_{clus} = \tau_{win,clus} - \tau_{loss,clus}.$$

## Unified Weighted Estimators & Estimands

These individual-pair and cluster-pair estimators and estimands can be unified within a general weighted framework, as described by Li and colleagues, which describes such a class of weighted average treatment effect estimands.[7] Let $\omega_{ij}$ denote a weight assigned to the comparison between clusters $i$ and $j$. The generalized weighted win probability estimator is:

$$\hat{\tau}_{win,\omega} = \frac{\sum_{i=1}^{M}\sum_{j=1}^{M} Z_i(1-Z_j)\omega_{ij}\frac{1}{N_i N_j}\sum_{k=1}^{N_i}\sum_{l=1}^{N_j} \mathrm{I}(Y_{ik} > Y_{jl})}{\sum_{i=1}^{M}\sum_{j=1}^{M} Z_i(1-Z_j)\omega_{ij}},$$

and the generalized weighted win probability estimand is then:

$$\tau_{win,\omega} = \frac{E\left[\omega_{ij}\frac{1}{N_i N_j}\sum_{k=1}^{N_i}\sum_{l=1}^{N_j} \mathrm{I}(Y_{ik}(1) > Y_{jl}(0))\right]}{E[\omega_{ij}]}.$$

Setting $\omega_{ij} = N_i N_j$ will then target the individual-pair estimand, whereas $\omega_{ij} = 1$ will target the cluster-pair estimand.

The weighted loss and tie probability estimators: $\hat{\tau}_{loss,\omega}$ and $\hat{\tau}_{tie,\omega}$, and estimands: $\tau_{loss,\omega}$ and $\tau_{tie,\omega}$, are analogously defined, as are the subsequent win estimators $(\widehat{WE}_{\omega})$: $\widehat{WR}_{\omega}$, $\widehat{WO}_{\omega}$, and $\widehat{WD}_{\omega}$, and win estimands $(WE_{\omega})$: $WR_{\omega}$, $WO_{\omega}$, and $WD_{\omega}$.

## Variance Estimation via Jackknife

For variance estimation and subsequent inference with the win estimators described above, we propose implementation of a leave-one-cluster-out jackknife variance estimator.[12] For each cluster $i \in \{1, \ldots, M\}$, we remove cluster $i$ and recalculate the estimates using the remaining $M - 1$ clusters. Deleting cluster $i$ removes all treatment-control cluster-pair comparisons involving cluster $i$. The resulting win probability estimator is then denoted as $\hat{\tau}_{win,\omega}^{(-i)}$, producing the resulting win estimators $\widehat{WE}_{\omega}^{(-i)}$ and jackknife variance estimator:

$$\widehat{Var}_{jk}(\widehat{WE}_\omega) = \frac{M-1}{M} \sum_{i=1}^{M} \left(\widehat{WE}_\omega^{(-i)} - \widehat{WE}_\omega\right)^2.$$

Statistical inference can then be robustly performed in CRTs using a t-distribution with $M - 2$ degrees of freedom.[13] Notably, related work has also suggested using $M - 1$ degrees of freedom.[7] The former $M - 2$ degrees of freedom will be more conservative with a small number of clusters $M$. Regardless, either degrees of freedom approach should generally yield comparable results, especially when there are a large number of clusters.

## When Do Individual-pair and Cluster-pair Estimands Differ?

When cluster sizes are uncorrelated with potential outcomes $\{Y_{ik}(0), Y_{ik}(1)\}$, then the previously described individual-pair and cluster-pair estimands will both be equivalent and reduce to:

$$\tau_{win,ind} = \tau_{win,clus} = P(Y_i(1) > Y_j(0)).$$

Without this correlation assumption, individual-pair and cluster-pair win estimands can substantially differ ($\tau_{win,ind} \neq \tau_{win,clus}$) when cluster sizes are informative (informative cluster size; ICS): that is, when (I.) baseline outcomes and/or (II.) treatment effects are dependent on cluster size. Previous work has found that, difference estimands typically only differ under type II ICS and are considered "collapsible".[6,14] In contrast, ratio estimands typically differ under both type I and/or II ICS, and are considered "non-collapsible".[6]

ICS can arise through several mechanisms. For example, there may be systematic differences in outcomes between small and large clusters (e.g., smaller hospitals may have different patient populations or care quality than larger hospitals). There may also be treatment effect heterogeneity by cluster size (e.g., interventions may be more effective in larger versus smaller clusters due to implementation factors). In general, greater variation in cluster size further amplifies the potential divergence between individual-pair and cluster-pair win estimands when cluster size is informative.

We illustrate the distinction between individual-pair and cluster-pair win estimands using hypothetical example CRTs with a single 3-level ordinal outcome $Y \in \{A, B, C\}$ (in descending order of preference) and informative cluster sizes. Appendix S2 includes an example with only type I ICS. In such a setting, individual-pair and cluster-pair estimands can lead to slightly different conclusions. However, these differences are modest and lead to largely similar interpretations, despite the large differences in sample sizes and baseline outcomes. This example illustrates that win estimands, even defined on the difference scale, are generally "non-collapsible" and can differ when outcomes are dependent on cluster size (Appendix S2). Notably, this defies some of the expectations set by Kahan and colleagues [6] regarding the collapsibility of difference estimands.

# Example: Ordinal outcome CRT with type I and II ICS

Here, we include a data example with type I and II ICS (Table 1). Suppose there are four cluster types ($T$), with different cluster sizes $N_T$ and equal proportions in the superpopulation of clusters ($P(T=1) = P(T=2) = P(T=3) = P(T=4) = 1/4$). With type I ICS, baseline outcomes (potential outcomes under control) differ by cluster size: larger cluster-types have higher proportions of outcome $C$'s (50% and 40% for clusters with $N_{T=1} = 1000$ and $N_{T=3} = 200$, respectively) compared to smaller cluster-types (25% for clusters with $N_{T=2} = 20$ and $N_{T=4} = 20$). With type II ICS, treatment effects vary by cluster size: larger cluster-types show large absolute percentage increases in outcome $A$ under treatment, while smaller cluster-types show large percentage decreases in outcome $A$ under treatment.

**Table 1. Example cluster randomized trial data with ordinal outcomes (type I and II ICS)**

| $T$ | $N_T$ | Under $Y(1)$ | | | Under $Y(0)$ | | |
|---|---|---|---|---|---|---|---|
| | | # of $A$'s | # of $B$'s | # of $C$'s | # of $A$'s | # of $B$'s | # of $C$'s |
| 1 | 1000 | 500 (50%) | 250 (25%) | 250 (25%) | 300 (30%) | 200 (20%) | 500 (50%) |
| 2 | 20 | 5 (25%) | 5 (25%) | 10 (50%) | 10 (50%) | 5 (25%) | 5 (25%) |
| 3 | 200 | 90 (45%) | 80 (40%) | 30 (15%) | 50 (25%) | 70 (35%) | 80 (40%) |
| 4 | 20 | 6 (30%) | 5 (25%) | 9 (45%) | 15 (75%) | 0 (0%) | 5 (25%) |

The win estimands corresponding to the CRT data example in Table 1 are calculated as described earlier and presented in Table 2. A separate example with more detailed estimand calculations is included in Appendix S2.

**Table 2. Example win estimands in a cluster randomized trial with ordinal outcomes (type I and II ICS)**

| Estimand | Win Ratio (WR) | Win Odds (WO) | Win Difference (WD) |
|---|---|---|---|
| Individual-pair | 2.238 | 1.700 | 0.259 |
| Cluster-pair | 0.880 | 0.920 | -0.042 |

In this example, the individual-pair and cluster-pair estimands lead to strikingly different conclusions. The individual-pair win estimands suggest strong treatment benefit (more preferred outcome), however the cluster-pair win estimands suggest treatment harm (fewer preferred outcome). This divergence arises because the individual-pair estimands are dominated by the two cluster-types with larger sizes ($N_{T=1} = 1000$ and $N_{T=3} = 200$), which contain a large majority of individuals and implies strong treatment benefit. In contrast, the cluster-pair estimand weighs all clusters equally, including those sampled from cluster-types with smaller sizes, and implies treatment harm. Altogether, this CRT scenario demonstrates that having both types of ICS can lead to dramatically different individual-pair and cluster-pair estimands.

# Simulation Study

We simulated data from a parallel CRT with $M = 100$ clusters, a single 5-level ordinal outcome, and potentially informative cluster sizes. Outcomes with and without type I and II ICS were generated from an underlying latent continuous outcome following a proportional odds model. Details regarding the simulation data generating process are included in Appendix S3.

The resulting individual-pair and cluster-pair win estimands in each simulation scenario ("no ICS" and "type I & II ICS") are then calculated via numerical integration and summarized in Table 3. As described in previous sections, the individual-pair and cluster-pair win estimands dramatically differ in the presence of such informative cluster sizes.

**Table 3. Simulation estimand values**

|   | Estimand | Individual-pair value | Cluster-pair value |
|---|---|---|---|
| **no ICS** | Win Ratio | 3.86 | 3.86 |
|  | Win Odds | 2.54 | 2.54 |
|  | Win Difference | 0.44 | 0.44 |
| **type I & II ICS** | Win Ratio | 1.90 | 1.13 |
|  | Win Odds | 1.49 | 1.07 |
|  | Win Difference | 0.20 | 0.03 |

The simulation results are summarized in Table 4. In scenarios with no ICS, both the individual-pair and cluster-pair win estimand results aligned, were unbiased, and yielded nominal coverage probabilities for the 95% confidence intervals (as formed using the jackknife variance estimator).

**Table 4. Simulation results**

|   | Estimand | Level | Relative Bias (%) | Coverage Probability |
|---|---|---|---|---|
| **no ICS** | WR | Individual | 2.6 | 0.954 |
|  | WO | Individual | 1.3 | 0.958 |
|  | WD | Individual | 0 | 0.952 |
|  | WR | Cluster | 2.4 | 0.953 |
|  | WO | Cluster | 1.2 | 0.952 |
|  | WD | Cluster | 0 | 0.955 |
| **ICS** | WR | Individual | 3.1 | 0.909 |
|  | WO | Individual | 0.4 | 0.910 |
|  | WD | Individual | 5.9 | 0.928 |
|  | WR | Cluster | 1.7 | 0.941 |
|  | WO | Cluster | 0.4 | 0.945 |
|  | WD | Cluster | 4.7 | 0.947 |

In scenarios with ICS, individual-pair estimators can exhibit some finite-sample bias, leading to poorer coverage and making them more challenging to target. However, even with ICS, the individual-pair estimators are still consistent and asymptotically unbiased by the weak law of large numbers, as highlighted earlier and further demonstrated by simulation in Appendix S4.

In contrast, the cluster-pair win estimator results in scenarios with type I and II ICS were unaffected by the finite-sample bias observed in individual-pair win estimands and yielded unbiased results with nominal coverage probabilities. Intuitively, cluster-pair ratio estimators assign equal weight to each cluster and only count each cluster pair once, avoiding the influence of larger clusters that drives the finite-sample bias.

# Discussion

Different treatment effect estimands based on win statistics, including the win ratio, win odds, and win difference are increasingly popular targets for summarizing endpoints in clinical trials.[1–3] These win estimands offer an intuitive approach for prioritizing outcomes by clinical importance. The implementation and interpretation of such win estimands can be complicated in cluster randomized trials (CRTs), where researchers can target fundamentally different estimands on the individual-level or cluster-level, with these estimands potentially leading to discordant conclusions.[5,6] In this article, we extend this CRT estimands-framework and clarify the distinction between individual-pair and cluster-pair win estimands in CRTs under the potential outcomes framework. We define the win, loss, and tie probability estimands as averages over the marginal distributions of the potential outcomes $Y(1)$ and $Y(0)$.[9–11] More discussion of the estimand definitions with potential outcomes are included in Appendix S1. We additionally introduce consistent estimators for each target estimand and propose variance estimation via convenient jackknife methods.

Altogether, we demonstrate through clear worked examples with ordinal outcomes that individual-pair and cluster-pair win estimands can differ when cluster size is informative and outcomes and/or treatment effects depend on cluster size. In settings with only type I ICS, where outcomes at baseline depend on cluster size, individual-pair and cluster-pair estimands can lead to slightly different conclusions (Appendix S2). However, these differences are modest and lead to largely the same interpretations, despite the large differences in sample sizes and baseline outcomes. Still, win estimands, even defined on the difference scale, are generally "non-collapsible" and can differ when outcomes are dependent on cluster size, defying some of the expectations set by Kahan and colleagues.[6] Subsequently, cases with both type I and II ICS, where outcomes and treatment effects both depend on cluster size, can produce individual-pair and cluster-pair win estimands with dramatically discordant results and even yield opposite conclusions regarding treatment benefit.

The dramatic differences between individual-pair and cluster-pair estimands is further illustrated by simulation with ordinal outcomes. Notably, our simulations highlight that despite demonstrated consistency, some caution should be taken when implementing individual-pair win estimators in the presence of ICS due to finite-sample bias. Unlike individual-pair estimators, cluster-pair estimators are generally unbiased for their respective targets.

We primarily focus on win estimands defined for ordinal outcomes to simply convey the impact of ICS on individual-pair and cluster-pair estimands in CRTs. Along with ordinal outcomes, the flexible non-parametric nature of win estimands have seen them applied for other complex outcome-types, including composite outcomes.[3,11,15] Although we do not explicitly highlight composite outcome examples in this article, the discussed discrepancies between differently weighted win estimands in the presence of informative cluster sizes are expected to extend to all outcome types.

This article presents individual-pair and cluster-pair win estimands under a unified general weighted estimand framework for CRTs. These differently weighted win estimands can differ in the presence of different types of informative cluster sizes in the simple parallel-CRT design, as highlighted in this article. Notably, more complicated multi-period CRT designs are common, including the parallel-with-baseline CRT (PB-CRT), cluster-randomized crossover design (CRXO), and stepped-wedge CRT (SW-CRT), for which more weighted estimands can be defined along with additional types of informative sizes.[16,17] Future work can further extend the general weighted estimand framework for win estimands to these more complicated CRT designs.

Careful specification of the target estimand is essential when applying win estimators in CRTs. The presence of informative cluster sizes can greatly complicate the analysis and interpretation of win estimators in cluster randomized trials. Failure to clearly define whether individual-pair or cluster-pair comparisons are of primary interest may result in answering a considerably different question than intended.

**Disclosure statement**

The authors declare no potential conflicts of interest with respect to the research, authorship, and/or publication of this article.

**Funding**

Research in this article was supported by the United States National Institutes of Health (NIH), National Heart, Lung, and Blood Institute (NHLBI, grant numbers 1R01HL178513 and 1R01HL168202).

**Disclaimer**

All statements in this report, including its findings and conclusions, are solely those of the authors and do not necessarily represent the views of the NIH. The authors declare that there are no conflicts of interest relevant to this work.

**ORCID**

Kenneth Menglin Lee (https://orcid.org/0000-0002-0454-4537)

Xi Fang (https://orcid.org/0000-0002-6692-5730)

Fan Li (https://orcid.org/0000-0001-6183-1893)

Michael O. Harhay (https://orcid.org/0000-0002-0553-674X)

**Data availability statement**

Data sharing is not applicable to this article as no new data were created or analyzed in this study. R code for the simulations are included in the online supplement.

# Appendix S1

We rely on the analog between win probability potential outcome estimands and the Mann-Whitney parameter in individual-randomized trials described in Mao[10] and Fay and colleagues[9], which defines the following win probability estimand:

$$\tau_{win} = P(Y_i(1) > Y_j(0)),$$

described in this present article, where $i$ can equal $j$. As per Mao[10] and Fay and colleagues[9], with a contrast function $h(t,s) = \mathrm{I}(t > s)$, comparing $t$ and $s$ with an indicator for $t > s$, it may initially seem intuitive to target $\tau_h^* = E[h(Y_i(1), Y_i(0))]$, which represents the individual causal effects averaged over the population. While $\tau_h^*$ resembles the above $\tau_{win}$, $\tau_h^*$ is in general not estimable because the joint distribution of $\{Y(1), Y(0)\}$ is not identifiable from observed data.[10] Fay and colleagues further discuss this, with attention given to Hand's paradox, where $\tau_h^*$ may differ from $\tau_{win}$.[9] Instead, Mao defines the population-level causal effect $\tau_{win}$ as an average over the marginal distributions, which are identifiable under standard assumptions:

$$\tau_{win} = \int \int \mathrm{I}(t > s) v_1(dt) v_0(ds) = P(Y_i(1) > Y_j(0))$$

where $v_Z$ is the marginal distribution of $Y(Z)$ with $Z = 0,1$.[10] With discrete outcomes, this estimand can be written as:

$$\tau_{win} = \sum_t \sum_s \mathrm{I}(t > s) P(Y(1) = t)) P(Y(0) = s).$$

Similarly, Fay and colleagues clearly lay out that the Mann-Whitney parameter estimand can be written as "one half plus the expected quantile difference between two groups".[9] Again, Fay and colleagues'[9] proposed estimand can then be demonstrated in to be equivalent to:

$$\tau_{win} = P(Y_i(1) > Y_j(0)).$$

Helpfully, Section 3.1 of their article includes a clear example of how the Mann-Whitney parameter estimand can be calculated with continuous potential outcomes $\{Y(1), Y(0)\}$,[9] resembling the ordinal examples in this present article.

To more clearly illustrate how the win estimands are calculated, we include calculation details for the individual-pair and cluster-pair potential outcome win estimands from the ordinal data example with type I informative cluster sizes in Appendix S2.

# Appendix S2

## Example: Ordinal outcome CRT with type I ICS

Suppose there are four cluster types ($T$), with different cluster sizes $N_T$ and equal proportions in the superpopulation of clusters ($P(T = 1) = P(T = 2) = P(T = 3) = P(T = 4) = 1/4$). Appendix Table S1 includes a data example with type I ICS, where baseline outcomes (potential outcomes under control) differ by cluster size: larger cluster-types have higher proportions of outcome $C$'s (50% and 40% for clusters with $N_{T=1} = 1000$ and $N_{T=3} = 200$, respectively) compared to smaller cluster-types (25% for clusters with $N_{T=2} = 20$ and $N_{T=4} = 20$). Overall, the treatment effects yield a constant decrease in outcome $C$ of 25%, increase in outcome $B$ of 5%, and increase in outcome of 20%. Notably, the following table extends the format of main manuscript Table 1 by explicitly including the individual marginal ("Ind Marginal Prob") and cluster marginal ("Cluster Marginal Prob") probabilities:

**Appendix Table S1. Example cluster randomized trial data with ordinal outcomes (type I ICS)**

| T | $N_T$ | Under $Y(1)$ | | | Under $Y(0)$ | | |
|---|---|---|---|---|---|---|---|
| | | # of $A$'s | # of $B$'s | # of $C$'s | # of $A$'s | # of $B$'s | # of $C$'s |
| 1 | 1000 | 500 (50%) | 250 (25%) | 250 (25%) | 300 (30%) | 200 (20%) | 500 (50%) |
| 2 | 20 | 14 (70%) | 6 (30%) | 0 (0%) | 10 (50%) | 5 (25%) | 5 (25%) |
| 3 | 200 | 90 (45%) | 80 (40%) | 30 (15%) | 50 (25%) | 70 (35%) | 80 (40%) |
| 4 | 20 | 19 (95%) | 1 (5%) | 0 (0%) | 15 (75%) | 0 (0%) | 5 (25%) |
| Individual Marginal Prob | | 623/1240 | 337/1240 | 280/1240 | 375/1240 | 275/1240 | 590/1240 |
| Cluster Marginal Prob | | 2.6/4 | 1/4 | 0.4/4 | 1.8/4 | 0.8/4 | 1.4/4 |

The win estimands corresponding to the CRT data example in Table 1 are calculated as described earlier, first with the individual-pair win probability estimand:

$$\tau_{win,ind} = \frac{E\left[\sum_{k=1}^{N_i} \sum_{l=1}^{N_j} I(Y_{ik}(1) > Y_{jl}(0))\right]}{E[N_i N_j]}.$$

The individual-pair win, loss, and tie probability estimands are then:

$$\tau_{win,ind} = \frac{623 \times (275 + 590) + 337 \times 590}{1240 \times 1240} = \frac{737,725}{1,537,600},$$

$$\tau_{loss,ind} = \frac{375 \times (337 + 280) + 275 \times 280}{1240 \times 1240} = \frac{308,375}{1,537,600},$$

$$\tau_{tie,ind} = \frac{623 \times 375 + 337 \times 275 + 280 \times 590}{1240 \times 1240} = \frac{491,500}{1,537,600}.$$

This yields the following individual-pair win estimands:

$$WR_{ind} = \frac{737,725}{308,375} \approx 2.392,$$

$$WO_{ind} = \frac{737,725 + 0.5 * 491,500}{308,375 + 0.5 * 491,500} \approx 1.775,$$

$$WD_{ind} = \left(\frac{737,725}{1,537,600}\right) - \left(\frac{308,375}{1,537,600}\right) \approx 0.279.$$

With the cluster-pair win probability estimand:

$$\tau_{win,clus} = E\left[\frac{1}{N_i N_j}\sum_{k=1}^{N_i}\sum_{l=1}^{N_j} I(Y_{ik}(1) > Y_{jl}(0))\right],$$

the cluster-pair win, loss, and tie probability estimands are then:

$$\tau_{win,clus} = \frac{2.6 \times (0.8 + 1.4) + 1 \times 1.4}{4 \times 4} = \frac{7.12}{16},$$

$$\tau_{loss,clus} = \frac{1.8 \times (1 + 0.4) + 0.8 \times 0.4}{4 \times 4} = \frac{2.84}{16},$$

$$\tau_{tie,clus} = \frac{2.6 \times 1.8 + 1 \times 0.8 + 0.4 \times 1.4}{4 \times 4} = \frac{6.04}{16}.$$

This yields the following cluster-pair win estimands:

$$WR_{clus} = \frac{7.12}{2.84} \approx 2.507,$$

$$WO_{clus} = \frac{7.12 + 0.5 * 6.04}{2.84 + 0.5 * 6.04} \approx 1.730,$$

$$WD_{clus} = \left(\frac{7.12}{16}\right) - \left(\frac{2.84}{16}\right) \approx 0.267.$$

With type I ICS, the individual-pair and cluster-pair estimands can lead to slightly different conclusions. However, these differences are modest and lead to largely the same interpretation, despite the large differences in sample sizes and baseline outcomes between the different cluster-types. Altogether, this example highlights that win estimands, even defined on the difference scale, are generally "non-collapsible" and can differ when outcomes are dependent on cluster size. Altogether, the potential non-collapsibility of the win difference defy some of the expectations set by Kahan and colleagues.[6]

## Appendix S3

We simulated data from a parallel CRT with 5-level ordinal outcomes summarized as win estimands, and potentially informative sizes. For individuals $k = 1, \ldots, N_i$ in clusters $i = 1, \ldots, M$, we simulate ordinal potential outcomes $Y_{ik}(1)$ and $Y_{ik}(0)$, corresponding to treatment and control. Outcomes with and without ICS are generated from an underlying latent continuous outcome following a proportional odds model. Conditional on cluster-type $T_i \in \{1, \ldots, \mathbb{T}\}$, latent cluster-level random intercept $\alpha_i$, and receiving treatment $z \in \{0,1\}$, the latent continuous potential outcome is:

$$Y_{ik}^*(z) = \mu_{T_i} + z\delta_{T_i} + \alpha_i$$

where $\mu_t$ is a cluster-type specific latent baseline value, $\delta_t$ is a cluster-type specific latent treatment effect, and latent cluster random intercept $\alpha_i \sim N(0, \tau_\alpha^2)$ independent of cluster-type $T_i$. Each cluster-type $T_i$ is independently drawn from a multinomial distribution with $P(T_i = t) = p_t$ which determines the latent baseline outcome distribution $\mu_t$, latent treatment effect $\delta_t$, and the expected cluster size $N_i$.

The ordinal potential outcomes $Y_{ik}(z) \in \{1, \ldots, D\}$ are obtained by first discretizing $Y_{ik}^*(z)$ using fixed cut-points:

$$-\infty = \theta_0 < \theta_1 < \cdots < \theta_{D-1} < \theta_D = \infty$$

such that $Y_{ik}(z) = r$ if $\theta_{r-1} < Y_{ik}^*(z) \leq \theta_r$. Equivalently, using the proportional odds model and conditional on $\alpha_i$, the cumulative probabilities satisfy:

$$P(Y_{ik}(z) \leq r \mid \alpha_i, T_i) = expit\left(\theta_r - \left(\mu_{T_i} + z\delta_{T_i} + \alpha_i\right)\right), r = 1, \ldots, D-1.$$

With the cluster-level SUTVA, the observed outcome for individual $k$ in cluster $i$ is then $Y_{ik} = Z_i Y_{ik}(1) + (1 - Z_i) Y_{ik}(0)$, where $Z_i \sim Bernoulli(0.5)$.

Cluster sizes are generated from a discrete uniform distribution that depends on the cluster-type $T_i$. Specifically, given integers $1 \leq L_{T_i} \leq U_{T_i}$:

$$N_i \sim Uniform(L_{T_i}, L_{T_i} + 1, \ldots, U_{T_i})$$

and

$$E[N_i] = \frac{L_{T_i} + U_{T_i}}{2}.$$

Altogether, this induces both type I and II ICS through the latent cluster-type $T_i$, which governs both the cluster size $(L_{T_i}, U_{T_i})$, latent baseline value $\mu_t$, and latent treatment effect $\delta_t$.

The simulation parameter values for scenarios without and with ICS are summarized in Table 3. In the scenarios without ICS: $p_t = \{0,1\}$ and $\gamma = 0$, such that there is only one cluster-type and no additional cluster size informativeness from latent cluster random intercept. In the scenarios with ICS: $p_t = \{0.92, 0.08\}$ and $\gamma = -0.6$, such that there are two cluster-types and additional cluster size informativeness from latent cluster random intercept.

**Appendix Table S2. Simulation parameter values**

| Parameter | Value | |
|---|---|---|
| Number of clusters ($M$) | 100 | |
| Number of cluster-types ($\mathbb{T}$) | 2 | |
| Ordinal categories ($D$) | 5 | |
| Cut-points ($\theta_r$) | $\{-1.2, -0.2, 0.6, 1.4\}$ | |
| Latent cluster intercept SD ($\tau_\alpha^2$) | 1.34 | |
| Monte-Carlo replicates | 2000 | |
| **Cluster-type specific values** | Value (No ICS) $\{t = 1, t = 2\}$ | Value (ICS) $\{t = 1, t = 2\}$ |
| Cluster-type probabilities ($p_t$) | $\{0, 1\}$ | $\{0.92, 0.08\}$ |
| Latent baseline values ($\mu_t$) | $\{2.2, 0.3\}$ | $\{2.2, 0.3\}$ |
| Treatment effects ($\delta_t$) | $\{0, 2.1\}$ | $\{0, 2.1\}$ |
| Cluster size ranges ($L_t, U_t$) | $\{8, 20\}$ | $\{80, 180\}$ |
| Cluster size average ($[L_t + U_t]/2$) | 14 | 130 |

# Appendix S4

Results from a single simulation replicate following the described data generating process (Appendix S3) with 100,000 clusters and informative sizes to demonstrate consistency and asymptotic unbiasedness of the individual-pair and cluster-pair win estimators, even in the presence of type I and II ICS.

**<u>Appendix Table S3.</u>**

|     | Estimand | Level | Relative Bias (%) |
|-----|----------|---------|------|
| ICS | WR | Individual | 1.4 |
|     | WO | Individual | 0.9 |
|     | WD | Individual | 2.1 |
|     | WR | Cluster | 0.0 |
|     | WO | Cluster | 0.0 |
|     | WD | Cluster | 0.5 |